1

# Using Graph-Pattern Association Rules On Yago Knowledge Base

**Wahyudi, Masayu Leylia Khodra, Ary Setijadi Prihatmanto, Carmadi Machbub**

School of Electrical Engineering and Informatics, Institut Teknologi Bandung
Bandung, Indonesia

wahyudi23509023@student.itb.ac.id, masayu@stei.itb.ac.id, asetijadi@lskk.ee.itb.ac.id, carmadi@stei.itb.ac.id

**Abstract.** We propose the use of Graph-Pattern Association Rules (GPARs) on the Yago knowledge base. Extending association rules for itemsets, GPARS can help to discover regularities between entities in knowledge bases. A rule-generated graph pattern (RGGP) algorithm was used for extracting rules from the Yago knowledge base and a graph-pattern association rules algorithm for creating association rules. Our research resulted in 1114 association rules, where the value of standard confidence at 50.18% was better than partial completeness assumption (PCA) confidence at 49.82%. Besides that the computation time for standard confidence was also better than for PCA confidence

## 1    Introduction

The Yago knowledge base [1] contains common sense knowledge. It is a collection of facts and information commonly known by humans. Yago was built by extraction of data from Wikipedia and using WordNet ontology. However, WordNet's ontology is limited, so Yago developed its own proprietary ontology. Yago evolved into Yago 2 [2] with the addition of data extracted from GeoNames, therefore Yago 2 can describe spatial entities using spatial data such as longitude and latitude from GeoNames. Yago 2 was expanded into Yago 3 [3] in view of the multilingual aspect. In previous versions Yago only extracted data from English Wikipedia, while in Yago 3 multilingual extraction from Wikipedia was done and grouping of entities was also done based on languages supported by Wikipedia.

Luis [4] developed the AMIE system to mine rules with incomplete facts using the approach of association rules [5]. AMIE uses the Yago KB and horn rules with data representation in the form of a relational database [6]. Luis explored horn rules and tuples contained in the relations of each entity. Each entity is represented by a function and has a maximum functionality value of 1 and a minimum functionality value of 0. The function has an inverse function, which also has a functionality value. For example: function *export(x,y)* has inverse function *isExported (y,x)*. Each functionality value of the function and the inverse function of each entity is compared and the largest value will be used by the system. The various connections between entities in the form of patterns was not discussed too much by Luis. The function and its inverse are used on the relational database to find rules. Partial completeness assumption (PCA) confidence is used to generate or predict negative evidence, but these measurements need more processing time and more computational resources than standard confidence. In contrast, our research focused on diversified graph pattern association to generate rules.

Fan [7] proposed association rules utilizing graph patterns. The proposed method uses parallel computation and focuses on social graphs. It obtains potential customers using the diversified mining problem (DMP) and entity identification problem (EIP) techniques based on the support of each entity. This is different for knowledge bases, especially when determining association rules. Fan claimed PCA confidence does not perform better than Bayes factor-based confidence if facts are represented in graph patterns. However, he used a different dataset under local closed world assumption (LCWA). Therefore we decided to use the graph pattern approach from Fan and the mining model from Luis. We wanted to investigate whether PCA confidence performs better than standard confidence when using graph patterns.

Our research used a combination of the techniques proposed by Luis [6] and Fan [7] on the Yago KB [8]. We used graph representation as data representation for connected data and for visualizing for the graph database. The flexibility of the graph model allows us to add entities and their relationships without affecting or modifying existing data [9]. Some well-known apps like Facebook, Google, Wikipedia and IMDB as well as many other apps use graph representation as data representation.

In this paper, we propose association rule mining of the Yago KB. More precisely, our contributions are: (1) we use graph-pattern association rules (GPARs) on the Yago knowledge base; (2) we define support and confidence for GPARs; 3) we experimentally verify the scalability and effectiveness of our algorithms for creating and mining rules.

The rest of this paper is structured as follows. Section 2 discusses related works. Section 3 introduces the preliminaries and Section 4 presents our mining model. Section 5 discusses the implementation of the graph-pattern association rules. Section 6 presents our experiments and Section 7 contains the conclusion and future works.

## 2      Related Work

Yago KB has the form of an RDF triple. RDF only has positive examples. It operates under open world assumption (OWA). This means that something not found in the KB is not necessarily assumed to be wrong but classified as unknown. This is a fundamental difference with database settings operating under closed world assumption (CWA). In CWA, facts that are not in the dataset cannot be assumed. For example, a KB contains the statement 'John was born in Paris'. Then there is the question: 'Was Alice also born in Paris?' Under CWA we get 'no' as the answer, while under OWA we get 'unknown' as the answer. CWA eliminates the possibility that Alice was born in Paris, while OWA keeps the possibility that Alice was born in Paris or not open.

Association rules were introduced by Agrawal [5]. Association rules combine multiple items into antecedents and have one item as consequent. Two steps are executed to generate association rules. Firstly, finding all itemsets that are present in at least c% of transactions. Secondly, finding association rules efficiently. Association rules have been well studied for discovering regularities between items in relational databases for promotional pricing and product placement. They have the traditional form of X ⇒ Y, where X and Y are disjoint itemsets.

Fanizzi [10] tried to mine rules from the semantic web using the inductive logic model (ILP). The goal was to find a hypothesis that included all positive examples in the absence of a negative example. This requires rules of various positive and negative examples to be investigated [11]. This is a problem in KBs because in KBs there are no negative examples. Another problem is that the ILP system cannot process large amounts of data while KBs contain a large amount of data.

Mining rules using ordinary techniques (inductive logic programming, logical rules) can only mine complete facts contained in a database. Incomplete facts cannot be used with this technique. Luis [12] used association rules under open world assumption (OWA) for KBs, introducing new thresholds for mining models called head coverage. This notion is used to filter rules based on size of the head, replacing the count of support as an absolute number. Moreover, it uses a new notion of confidence measurement called partial completeness assumption (PCA) confidence. Our research applied this confidence for comparison with standard confidence [5].

The graph-pattern association rules proposed by Fan [7] were used to create graph patterns for mining association rules in social media marketing and identifying potential customers, using parallel computation. There are existing algorithms for pattern mining graph databases. Large-scale mining techniques in a single graph have also been studied [13], notably top-k algorithms to reduce cost and scalable subgraph isomorphism algorithms adapted to generate pattern candidates [13].

Yago knowledge base graph properties [14] can be seen as a set of facts, where each fact consists of two nodes that are connected by one edge *(x,r,y)* with *x* denoting node 1, *r* the relation (or edge), and *y* denoting node 2 of the fact. There are several equivalent alternative representations of facts. In this study, we borrowed the notation from Datalog and represent facts as *r(x, y)*. For example, we write *isLocatedIn(Bandung,Indonesia)*.

## 3    Mining Model

In this paper, we focus on Yago knowledge base graph properties [14]. A graph property model is a graph consisting of nodes, edge, and properties [9]. We use properties such as entity properties. Each node has properties, as depicted in Figure 1. It shows two nodes with the person label and a node with the book label. The two nodes are connected with the edge label *hasRead*. The person node has the property name and value *John Smith*, and the book node has two sets of properties, *title* and *author*. The title has the value *graph database* and the author has the value *Ian Robinson*.

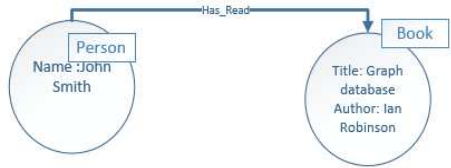

**Figure 1**  Graph property model

In this section, we will explain the mining model that will be generated. Most of the models we present are adopted from Luis [15] and Fan [7]. We adopted the approach of Luis for support, head coverage and confidence for graph patterns. We used a different approach than Fan's for graph patterns (see Section 4). The difference is explained as follows:

**Support**. The support of a rule quantifies the number of correct rules, i.e. the size of A. A rule's support is the frequency or number of itemsets in the data set. Support is calculated from the calculated number of itemsets compared to the total number of itemsets in the dataset. Support graph pattern *P* in a graph *G*, denoted by *supp(P,G)*, indicates the number of *Ps* contained in *G*. Our approach uses support as the number of instantiations of a rule that appear in a KB. The support of graph pattern *p* is denoted by *supp(P (G))*. This is the number of nodes and edge pairs found in graph pattern *P (G)*. Support of rule *R* is denoted by *supp R*, i.e. the number of nodes and edge pairs present in *R*.

$$supp(P_{(G)}) = |P_{(G)}|, \ supp\ R = |R|$$

**Head Coverage**. As mentioned above, we use head coverage (HC) as threshold for the strength of a rule. In association rules usually *min support* is used as the threshold for the strength of a rule. In this research, we use HC as threshold for the strength of a rule. We used *min HC =* 0.01.

$$R: P(G) \Rightarrow r(u, w) \tag{1}$$

$$HC = \frac{supp\,(R)}{|r|} \tag{2}$$

where *HC* is head coverage, *supp (R)* is the number of rules *R*, and | r | is the size of the head in the dataset.

**Confidence.** This is a measure to determine the strength of a rule. The value is between 0 and 1. A rule with high confidence is close to 1 and, vice versa, a rule with low confidence is close to 0. In this research, we use two types of confidence: standard confidence and PCA confidence.

**Standard Confidence**. Standard confidence (*conf*) is a measure of the ratio of the number of rules *R* compared to the facts we know in the form of graph pattern *P (G)*, as in the equation below:

$$conf(R_i) = \frac{|R_i|}{|P_{(G)i}|} \tag{3}$$

**PCA Confidence**. Standard confidence does not distinguish between facts that are not in the dataset and wrong facts in the dataset. In other words, standard confidence cannot distinguish between wrong facts and unknown facts. Since the knowledge base has no negative facts, the partial completeness approach (PCA) was used, as proposed by Luis [12]. If *r (u, w)* ∈ *G* for nodes *u* and *w* then:

$$\forall w' = r(u, w) \in G \cup new\ true \Rightarrow r(u, w') \in G$$

In other words, we assume that if graph *G* knows some attribute *x* of *u*, then it all attributes *x* of *u* can be seen. This assumption is converted to standard confidence, so the following is obtained:

$$PCA\ conf(R_i) = \frac{|R_i|}{|P(G)_i \wedge r(u,w')|} \tag{4}$$

## 4    Graph-pattern Association Rules

In this section we will discuss the approach used in this research in detail: graph-pattern association rule $\mathcal{P}(x, y)$ is defined as $B(x, y) \Rightarrow r(x, y)$, where $B(x, y)$ is a graph pattern in which $x$ and $y$ are two designated nodes and $r(x, y)$ is an edge labeled $r$ from $x$ to $y$, on which the same search conditions as in $B$ are imposed. We refer to $B$ and $r$ as the antecedent and consequent of $\mathcal{P}$, respectively [7].

We use graph patterns to mine association rules. We chose this option for the following reasons: we follow the Agrawal model by starting from 1 antecedent and increasing the antecedents to 2, 3 and 4. In the tool we use, we can automatically search for a graph pattern, but there is a problem with the edge direction: there is only one edge direction. This is certainly different from the actual data in the database, where the relationship can come from two sides, from the subject or the object. Figure 2 shows two graph patterns that have different directions on R2. In Figure 2a, R2 has the same direction as R1 but in Figure 2b, R2 has the opposite direction from R1. This cannot be executed by one query, instead there should be a query for each graph pattern. In graph theory, the graph motif technique is used, i.e. the isomorphism of two subgraphs is determined by the interaction pattern between node and edge. In the proposed method, we use the edge property (relation) as the motif for determining the isomorphism of subgraphs. Therefore we use the graph pattern to get all possible directions from each of the existing entities. This is different from Fan [7], who used a pattern generator [13].

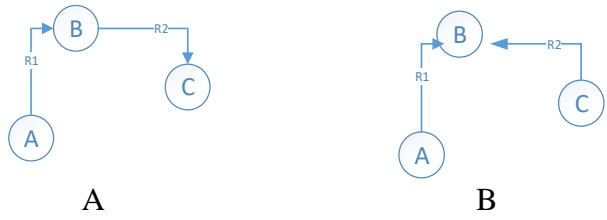

        A                              B

**Figure 2** Graph Pattern

We use ten graph patterns, consisting of patterns that have one relation, two relations and, three relations. Each graph pattern has 2 types of consequents. Twenty graph patterns were used, as shown in Table 1 below.

**Table 1** Graph Patterns used in this Research

| No | Graph Pattern | Association rule | Association rule graph pattern |
|----|---------------|------------------|--------------------------------|
| 1 | | $A - [r1] \rightarrow B => B - [r2] \rightarrow A$<br>$A - [r1] \rightarrow B => A - [r2] \rightarrow B$ | $A - [r1] \rightarrow B - [r2] \rightarrow A$<br>$A - [r1] \rightarrow B \leftarrow [r2] - A$ |
| 2 | | $A - [r1] \rightarrow B - [r2] \rightarrow C => A - [r3] \rightarrow C$<br>$A - [r1] \rightarrow B - [r2] \rightarrow C => C - [r3] \rightarrow A$ | $A - [r1] \rightarrow B - [r2] \rightarrow C - [r3] - A$<br>$A - [r1] \rightarrow B - [r2] \rightarrow C - [r3] \rightarrow A$ |
| 3 | | $A - [r1] \rightarrow B \leftarrow [r2] - C => A - [r3] \rightarrow C$<br>$A - [r1] \rightarrow B \leftarrow [r2] - C => C - [r3] \rightarrow A$ | $A - [r1] \rightarrow B \leftarrow [r2] - C - [r3] - A$<br>$A - [r1] \rightarrow B \leftarrow [r2] - C - [r3] \rightarrow A$ |
| 4 | | $A \leftarrow [r1] - B - [r2] \rightarrow C => A - [r3] \rightarrow C$<br>$A \leftarrow [r1] - B - [r2] \rightarrow C => C - [r3] \rightarrow A$ | $A \leftarrow [r1] - B - [r2] \rightarrow C - [r3] - A$<br>$A \leftarrow [r1] - B - [r2] \rightarrow C - [r3] \rightarrow A$ |
| 5 | | $A - [r1] \rightarrow B - [r2] \rightarrow C - [r3] \rightarrow D => A - [r4] \rightarrow D$<br>$A - [r1] \rightarrow B - [r2] \rightarrow C - [r3] \rightarrow D => D - [r4] \rightarrow A$ | $A - [r1] \rightarrow B - [r2] \rightarrow C - [r3] \rightarrow D \leftarrow [r4] - A$<br>$A - [r1] \rightarrow B - [r2] \rightarrow C - [r3] \rightarrow D - [r4] \rightarrow A$ |
| 6 | | $A - [r1] \rightarrow B - [r2] \rightarrow C \leftarrow [r3] - D => A - [r4] \rightarrow D$<br>$A - [r1] \rightarrow B - [r2] \rightarrow C \leftarrow [r3] - D => D - [r4] \rightarrow A$ | $A - [r1] \rightarrow B - [r2] \rightarrow C \leftarrow [r3] - D \leftarrow [r4] - A$<br>$A - [r1] \rightarrow B - [r2] \rightarrow C \leftarrow [r3] - D - [r4] \rightarrow A$ |
| 7 | | $A \leftarrow [r1] - B - [r2] \rightarrow C - [r3] \rightarrow D => A - [r4] \rightarrow D$<br>$A \leftarrow [r1] - B - [r2] \rightarrow C - [r3] \rightarrow D => D - [r4] \rightarrow A$ | $A \leftarrow [r1] - B - [r2] \rightarrow C - [r3] \rightarrow D \leftarrow [r4] - A$<br>$A \leftarrow [r1] - B - [r2] \rightarrow C - [r3] \rightarrow D - [r4] \rightarrow A$ |
| 8 | | $A \leftarrow [r1] - B - [r2] \rightarrow C \leftarrow [r3] - D => A - [r4] \rightarrow D$<br>$A \leftarrow [r1] - B - [r2] \rightarrow C \leftarrow [r3] - D => D - [r4] \rightarrow A$ | $A \leftarrow [r1] - B - [r2] \rightarrow C \leftarrow [r3] - D \leftarrow [r4] - A$<br>$A \leftarrow [r1] - B - [r2] \rightarrow C \leftarrow [r3] - D - [r4] \rightarrow A$ |
| 9 | | $A - [r1] \rightarrow B \leftarrow [r2] - C - [r3] \rightarrow D => A - [r4] \rightarrow D$<br>$A - [r1] \rightarrow B \leftarrow [r2] - C - [r3] \rightarrow D => D - [r4] \rightarrow A$ | $A - [r1] \rightarrow B \leftarrow [r2] - C - [r3] \rightarrow D \leftarrow [r4] - A$<br>$A - [r1] \rightarrow B \leftarrow [r2] - C - [r3] \rightarrow D - [r4] \rightarrow A$ |
| 10 | | $A - [r1] \rightarrow B \leftarrow [r2] - C \leftarrow [r3] - D => A - [r4] \rightarrow D$<br>$A - [r1] \rightarrow B \leftarrow [r2] - C \leftarrow [r3] - D => D - [r4] \rightarrow A$ | $A - [r1] \rightarrow B \leftarrow [r2] - C \leftarrow [r3] - D \leftarrow [r4] - A$<br>$A - [r1] \rightarrow B \leftarrow [r2] - C \leftarrow [r3] - D - [r4] \rightarrow A$ |

We developed the algorithm shown in Figure 3 to generate rules from a graph pattern. This algorithm is called the rule-generated graph pattern (RGGP) algorithm. It has as input a knowledge base and graph patterns. In the first step, e sort the twenty graph patterns in an array and select them one by one with a loop. The expected results are body rules, rule heads and the number of rules generated.

```
function generateRule (KB K, Graph pattern Gp)
gp = [p1,p2....pn]
create collection
for (i=1;i<=n;i++) do
 rb= pi.extractbody();
 rh= pi.extracthead();
 c=count(pi)
 collection.add (<rb,rh,c>)
 end for
return collection
```

**Figure 3**  Rule-generated graph pattern (RGGP) algorithm.

Finally, rules are generated from the RGGP algorithm. The next step is to determine the rules to be used in the association rules. The algorithm used is the graph-pattern association rule (GPARS) algorithm shown in Figure 4. This algorithm has input in the form of the collection rules that were generated by the RGGP algorithm. The first step of the algorithm is to process the rules one by one, from the first rule to the last, after the values of head coverage, support, standard confidence, and PCA confidence have been set to zero. Each rule counts the number of body rules and head rules, after which the value of head coverage is counted. Rules that qualify are rules that have minimum head coverage >= 0.01. Support, standard confidence and PCA confidence are calculated for each rule.

```
function argp (collection,
MinHC,supp,stdconf,pcaconf)
out ={}
minHC = 0
supp=0
stdconf = 0
pcaconf = 0
for (i=0;i<=n;i++) do
  r=collection.get(i).c
  rhead = collection.get(i).rh
  rbody = collection.get(i).rb
  ch = count(rhead)
  cb = count (rbody)
  for all instantiations rhead(x, y)replace rhead
  by rpca, y by z
  minHC = r /ch
  if (minHC >=0.001) then
   supp r = count r;
   cpca = count rpca
   stdconf = supp r / cb
   pcaconf = supp r / cpca
   out.add{r,minHC,supp r,stdconf,pcaconf}
  end if
end for
return out
```

**Figure 4** Graph-pattern association rule (GPARS) algorithm.

## 5    Experiment

Using the graph properties of the Yago KB, we conducted an experiment to generate collection rules and association rules. The experiment used the 20 types of graph patterns shown in Table 1. It also used standard confidence and PCA confidence, because we wanted to investigate whether PCA confidence performs better than standard confidence when using graph patterns.

### 5.1    Experimental setup

We used Neo4j for visualizing the graph database and graph processing. Our dataset, Yago, had 600 Kb nodes of more than 50 different types, and 980 Kb edges of 30 types, such as *isPoliticianOf, isLeaderOf*, etc. All experiments ran on a laptop with 8 GB of RAM and four physical CPUs (Intel core i3 at 1.7 GHz).

## 5.2 Standard Confidence vs. PCA Confidence

This experiment generated 1114 rules that met head coverage >= 0.01. For each rule its confidence was calculated using standard confidence and PCA confidence. From the result of the experiment 559 rules had standard confidence better than PCA confidence (50.18%), whereas 555 rules (49.82%) had PCA confidence better than standard confidence. Table 2 below shows the 3 rules that had the best standard confidence vs. PCA confidence.

**Table 2**    Top 3 Rules for Standard Confidence vs. PCA Confidence

| No | Graph pattern | Standard Confidence | PCA Confidence |
|---|---|---|---|
| 1 | a<-[isPoliticianOf]-b-[isLeaderOf]->c<-[isLeaderOf]-d => d-[isPoliticianOf]->a | 0.86 | 1. |
| 2 | a<-[isMarriedTo]-b-[hasChild]->c => a-[hasChild]->c | 0.59 | 0.74 |
| 3 | a-[isLeaderOf]->b-[diedIn]->c-[isLocatedIn]->d => a-[isPoliticianOf]->d | 0.53 | 0.018 |

| No | Graph pattern | Standard Confidence | PCA Confidence |
|---|---|---|---|
| 1 | a<-[isPoliticianOf]-b-[isLeaderOf]->c<-[isLeaderOf]-d => d-[isPoliticianOf]->a | 0.86 | 1 |
| 2 | a-[hasChild]->b<-[hasChild]-c => c-[isMarriedTo]->a | 0.586 | 0.738 |
| 3 | a<-[isMarriedTo]-b-[hasChild]->c => a-[hasChild]->c | 0.589 | 0.738 |

## 6    Conclusion and Future Works

In this paper, graph-pattern association rules (GPARs) were proposed for itemsets in syntax and semantics to support confidence metrics and graph properties for mining association rules from the Yago knowledge base. Our confidence metrics used standard confidence and PCA confidence. The experimental result indicated that standard confidence performed slightly better than PCA confidence. We obtained an average value for PCA confidence that was lower than that of standard confidence for the graph patterns. Therefore, further research is expected to determine the appropriate confidence for graph patterns.